\documentclass[12pt]{spieman}  
\usepackage{amsmath,amsfonts,amssymb}
\usepackage{graphicx}
\usepackage{setspace}
\usepackage{tocloft}
\usepackage{enumitem}
\usepackage{booktabs} 
\usepackage{comment}
\usepackage{gensymb}
\usepackage{textcomp}

\usepackage{lineno}

\title{Random Vibration Testing of Microelectromechanical Deformable Mirrors for Space-based High-Contrast Imaging}

\author[a,*]{Axel Potier}
\author[a]{Camilo Mejia Prada}
\author[a]{Garreth Ruane}
\author[a]{Hong Tang}
\author[a]{Wesley Baxter}
\author[a]{Duncan Liu}
\author[a]{A J Eldorado Riggs}
\author[a]{Phillip K. Poon}
\author[a]{Eduardo Bendek}
\author[a]{Nick Siegler}
\author[a]{Mary Soria}
\author[a]{Mark Hetzel}
\author[b]{Charlie Lam}
\author[b]{Paul Bierden}
\affil[a]{Jet Propulsion Laboratory, California Institute of Technology, 4800 Oak Grove Drive,Pasadena, CA 91109}
\affil[b]{Boston Micromachines Corp, 30 Spinelli Place, Cambridge, MA 02138, USA}

\cftpagenumbersoff{figure}
\cftpagenumbersoff{table} 
\begin{document} 
\maketitle

\begin{abstract}
Space-based stellar coronagraph instruments aim to directly image exoplanets that are a fraction of an arcsecond separation and ten billion times fainter than their host star. To achieve this, one or more deformable mirrors (DMs) are used in concert with coronagraph masks to control the wavefront and minimize diffracted starlight in a region of the image known as the ``dark zone" or ``dark hole."  The DMs must have a high number of actuators (50 to 96 across) to allow dark holes that are large enough to image a range of desired exoplanet separations. In addition, the surfaces of the DMs must be controlled at the picometer level to enable the required contrast. Any defect in the mechanical structure of the DMs or electronic system could significantly impact the scientific potential of the mission. Thus, NASA's Exoplanet Exploration Program (ExEP) procured two 50$\times$50 microelectromechanical (MEMS) DMs manufactured by Boston Micromachines Corporation (BMC) to test their robustness to the vibrational environment that the DMs will be exposed to during launch. The DMs were subjected to a battery of functional and high-contrast imaging tests before and after exposure to flight-like random vibrations. The DMs did not show any significant functional nor performance degradation at $10^{-8}$ contrast levels. 
\end{abstract}

\keywords{Deformable mirrors, High-contrast imaging, Wavefront sensing and control, Exoplanets}

{\noindent \footnotesize\textbf{*}Axel Potier,  \linkable{axel.q.potier@jpl.nasa,gov} }

\begin{spacing}{2}   

\section{Introduction}
\label{sec:introduction}  

The 2020 Decadal Survey on Astronomy and Astrophysics\cite{Astro2020} prioritized the development of technologies for directly imaging Earth-like exoplanets with the future Habitable World Observatory (HWO) flagship mission. The document has recommended a large ($\sim$6-meter) infrared/ optical/ ultraviolet (IR/O/UV) telescope with a stellar coronagraph or starshade to be launched in the first half of the 2040's. If a coronagraph instrument is selected, it will be designed to attenuate the diffracted light from the host star to create a region of high contrast in the image (known as the ``dark zone" or ``dark hole") where exoplanets that are $\sim10^{10}$ times fainter than the star may be imaged at angular separations of $<$1 arcsecond. To accomplish this, a series of coronagraph masks and one or more deformable mirrors (DMs) will be used to minimize the stellar intensity in the dark hole\cite{Malbet1995,Trauger2004}. 

DM technologies are being developed for both ground- and space-based high-contrast imaging applications. Electrostrictive devices\cite{Ealey2004,Wirth2013} and microelectromechanical (MEMS)\cite{Bifano2011,Morgan2019,Bendek2020} systems are the most advanced for space applications and have been at least partially flight qualified. On one hand, the Roman Space Telescope (RST) Coronagraph Instrument will make use of two 48$\times$48 electrostrictive devices manufactured by AOA Xinetics\cite{Kasdin2020}. These DMs have proven high reliability and high performance but the contact between the electrodes and the reflective face sheet may increase the potential for unwanted motions due to thermal variations. Moreover, these devices have a relatively large ($\ge$1~mm) inter-actuator pitch. On the other hand, MEMS DMs have also demonstrated promising performance\cite{Baudoz2018SPIE,MejiaPrada2019,Riggs2021}. Their contactless technology mitigates hysteresis and other instabilities caused by environmental factors and allows a smaller pitch (0.3-0.4~mm), which makes them attractive candidates for the future HWO mission\cite{HabEx_finalReport,LUVOIR_finalReport}. However, the technological readiness of the MEMS DMs lags behind the electrostrictive DMs that will be demonstrated in flight by RST.  Indeed, lead magnesium niobate electroceramic actuators (PMN), manufactured by AOA Xinetics, have successfully completed the full space qualification process for use in RST Coronagraph Instrument. However, the mechanical construction differs, making it impossible to transfer the heritage to MEMS devices.

Boston Micromachines Corporation (BMC) MEMS DMs have been extensively tested in vacuum at High Contrast Imaging Testbed (HCIT) demonstrating their ability to function and endure in a vacuum. However, it was found that the absence of air allows residual high-frequency electrical noise to cause mechanical resonance of the DM membrane. This issue was resolved by implementing RC filters on each channel to dampen the electrical noise before it could induce mechanical vibration\cite{Bendek2020}. Thereafter, proper operation in a vacuum chamber was successfully demonstrated\cite{Riggs2021}.

The next milestone in space qualification for the MEMS DM is proving its ability to survive the General Environmental Verification Standard (GEVS) vibration profile. But, we inform the reader that assessing the survivability and operational capabilities of DMs during launch and in space would require multiple additional stages. Before allocating more resources for further tests, such as acoustics, shock, radiation, and EMI, we first sought to confirm the MEMS DM's endurance under these conditions. Prior attempts to evaluate MEMS vibration survivability revealed that the tested devices exhibited altered behavior following the shake and vibe process\cite{Bierden2022}. However, after discussing the experimental setup with the authors, we concluded that the change might have been caused by other factors, such as particle contamination. Lacking sufficient information to evaluate the MEMS DM's vibration resistance, we decided to conduct a new study, which is presented in this paper. 

In that context, and as part of a NASA Small Business Innovation Research project (SBIR) titled ``Improved Yield, Performance and Reliability of High-Actuator-Count Deformable Mirrors", BMC developed a new fabrication process and several design modifications that were integrated in a complete fabrication cycle, producing fully operational 2040-actuator continuous face-sheet MEMS DMs.

With support from NASA's Exoplanet Exploration Program’s (ExEP) coronagraph technology development efforts, several sets of these DMs were tested for reliability in a flight-like environment. Expanding on our previous results\cite{MejiaPrada2021}, this paper reports on experiments carried out at the HCIT facility at NASA's Jet Propulsion Laboratory to demonstrate the robustness of BMC's MEMS DM technology to random vibrations during rocket launch. In Sec.~\ref{sec:Description}, we present the manufacturing and integration of these DMs and their design from the dedicated electronics to the DM face-sheet. In Sec.~\ref{sec:Environmental_Testing}, we describe the workflow of the tests performed in HCIT and the different facilities used in these experiments. Finally, Sec.~\ref{sec:Results} presents the results that demonstrate the survivability and robustness of high-actuator-count MEMS DMs to random vibrations.

\section{Design and fabrication of the MEMS DMs}
\label{sec:Description}

\begin{figure}[t]
    \centering
	\includegraphics[width=0.8\linewidth]{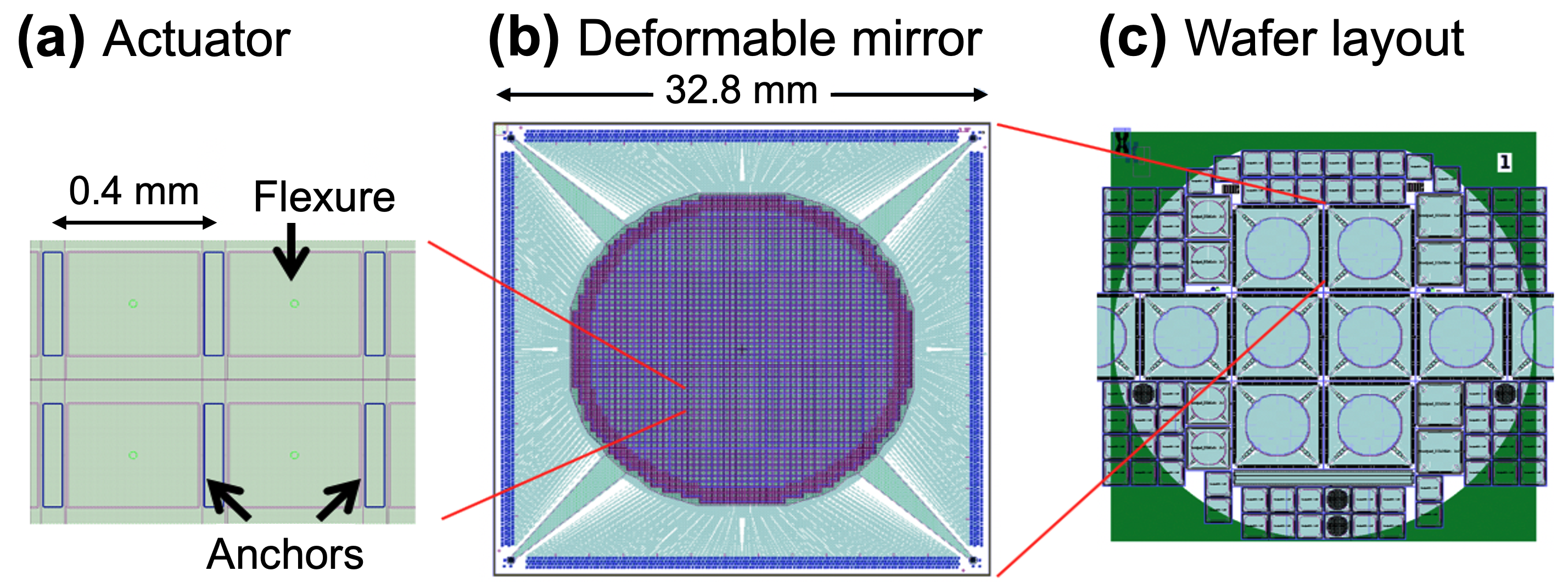}
	\caption{Diagram of the devices under test. (a) A zoomed-in view of a 2$\times$2 actuator region. Each actuator (0.4~mm across) consists of a flexure anchored at its edges. (b) The full DM layout with 2040 actuators as well as wire traces extending radially to the edges of the 32.8~mm substrate. (c) The full wafer layout as manufactured, which typically has several DMs of different formats to make best use of the available wafer area.} 
	\label{fig:MEMS_design} 
\end{figure}

\subsection{DM design and wafer fabrication}
For this study, we procured two 2040-actuator DMs from BMC with the characteristics specified in Table~\ref{table:MEMS_properties} and layout illustrated in Fig.~\ref{fig:MEMS_design}. The DMs were manufactured on 1.1~mm thick substrates that is more than four times stiffer than the standard substrate. This thickness was optimized through a finite element analysis studying the stresses on the mirror caused by random vibrations  using conservative RST coronagraph instrument flight acceptance specifications (see blue curve in Fig.~\ref{fig:PSD_Vibrations}). It brings about a higher resistance to bending stresses exerted by the thin films. This change also reduces the surface deformation of the unpowered DM and then increases the DM usable stroke for wavefront control after flattening. BMC developed custom tooling at the commercial MEMS foundry to work with the new substrate and process these thicker wafers.

\begin{table}[htp]
	\caption{High-actuator-count MEMS DMs properties}
	\centering
	\begin{tabular}{lc}
		\toprule
		\# of actuators in active area & 2040 \\
		\midrule
		 \# of actuators across the active area & 50 \\
		 \midrule
		 Pupil diameter & 19.6~mm \\
		 \midrule
		 Actuator pitch & 400~$\mu $m \\
		 \midrule
		 Substrate thickness & 1.1~mm \\
		 \midrule
		 Actuator stroke & 1.0~$\mu $m \\
		 \midrule
		 Operating voltage &  0-98V\\
		 \midrule
		\bottomrule
	\end{tabular}
	\label{table:MEMS_properties}
\end{table}

Beside the substrate thickness, no step in the manufacturing process is different from usual BMC DMs. The procedure, illustrated in Fig.~\ref{fig:MEMS_Manufacturing}, was as follows:

\begin{figure}[t]
    \centering
	\includegraphics[width=\linewidth]{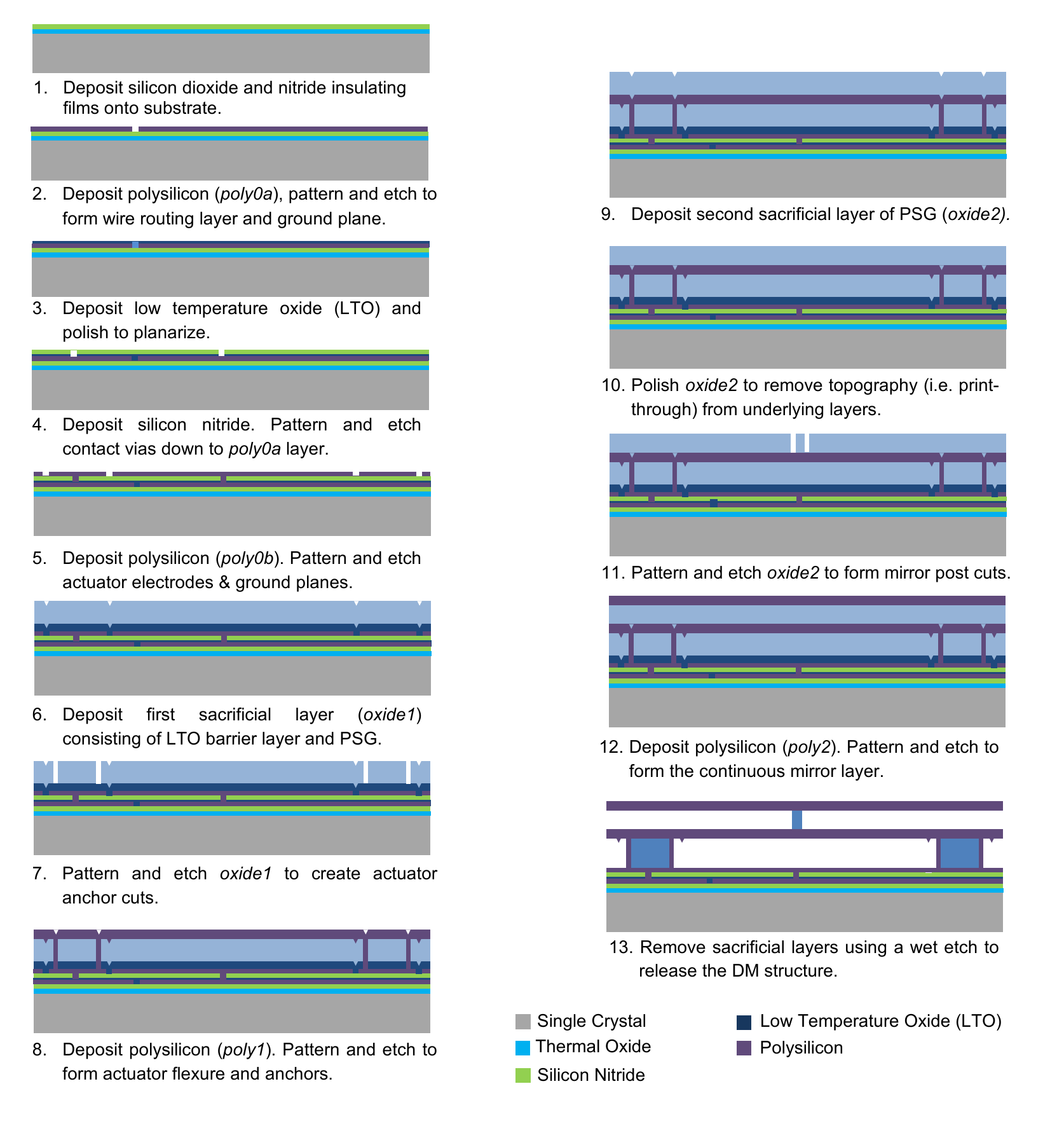}
	\caption{BMC's MEMS DM fabrication process.} 
	\label{fig:MEMS_Manufacturing} 
\end{figure}

\begin{enumerate}[label=(\roman*)]
\item Silicon dioxide and a low stress silicon nitride layer were deposited on a single crystal silicon substrate to electrically insulate the conductive substrate from the MEMS devices (Fig.~\ref{fig:MEMS_Manufacturing}, step 1).
\item The first layer of polysilicon (referred to as Poly 0a) was then deposited, patterned, and etched to create actuator base electrodes and wire routing for the array (Fig.~\ref{fig:MEMS_Manufacturing}, step 2).
\item A low-temperature oxide (LTO) layer was deposited and polished using chemo-mechanical polishing techniques to flatten the layer. A second dielectric film, low-stress silicon nitride was then deposited (Fig.~\ref{fig:MEMS_Manufacturing}, step 3-4).
\item The LTO and silicon nitride layer was lithographically patterned and etched to provide a path for electrical connectivity between the wire traces, actuator electrodes, and grounded landing pads, which were
produced in a subsequent polysilicon thin film (Poly 0b) deposition and patterning process (Fig.~\ref{fig:MEMS_Manufacturing}, steps 4).
\item An array of actuator electrodes was fabricated (Fig.~\ref{fig:MEMS_Manufacturing}, step 5). Then, a thick sacrificial layer (Oxide 1) made up of phosphosilicate glass (PSG) and a thin barrier layer of LTO was deposited on the Poly 0b layer to create the actuator gap (Fig.~\ref{fig:MEMS_Manufacturing}, step 6). The stroke of the electrostatic actuators depends on its thickness.
\item The Oxide 1 film was patterned and etched once more to create the actuator anchor features (Fig.~\ref{fig:MEMS_Manufacturing}, step 7).
\item A second layer of polysilicon, Poly 1, was deposited, patterned, and etched to create actuator anchors and compliant actuator flexures with integrated hard stops (Fig.~\ref{fig:MEMS_Manufacturing}, step 8).
\item A second sacrificial layer, Oxide 2, was deposited and polished to remove print-through from the underlying films. The mirror attachment post features were then patterned and etched into the Oxide 2 film (Fig.~\ref{fig:MEMS_Manufacturing}, steps 9-11).
\item A final polysilicon layer, Poly 2, was then deposited, polished, and patterned (Fig.~\ref{fig:MEMS_Manufacturing}, step 12).
\item Metal bond-pads were added to allow for wire-bonding of the device, and to allow the wafers to be diced into individual DM devices.
\item Finally, the sacrificial oxide layers were removed using a wet etch “release” process (Fig.~\ref{fig:MEMS_Manufacturing}, step 13).
\end{enumerate}

Once the devices were received from the foundry, BMC inspected the DMs using visible and IR optical microscopy and interferometry to identify potential optical, electrical, and subsurface manufacturing defects. Using a custom probe station, each candidate die that passed this initial inspection was tested to determine actuator response. Their electro-mechanical and optical performance were then characterized, including responsiveness of each actuator, stroke limits, unpowered surface error, and actuator defects. 

We selected two devices for this study, which we refer to as Device Under Test (DUT)~1 and~2. Their initial surface properties are summarized in Table~\ref{table:Surface_properties}. DUT 1 is a 100\% functional unit that was coated with an evaporated thin film of aluminum at BMC’s facility. The purpose of DUT~1 is to confirm or reject the hypothesis that a fully functional 2K MEMS DM can survive a launch environment. On the other hand, DUT 2 had some unresponsive actuators. We kept its face sheet uncoated to allow for the post vibration infrared inspection of the DM surface to help understanding any failure mode. This DUT aimed to test the hypothesis that defects causing anomalous actuators can propagate to neighboring actuators during random vibrations\cite{MejiaPrada2021}.

\begin{table}[t]
	\caption{Initial surface quality properties of DUT~1 and DUT~2. Unpowered and powered surface errors are domniated by a strong astigmatism, characteristic of MEMS DMs.}
	\centering
	\begin{tabular}{lcc}
		\toprule
		 & DUT 1 & DUT 2\\
		 \midrule
		 Initial yield & 100\% & 99.3\% \\
		\midrule
		 Unpowered surface deformation (PV/RMS, in nm) & 604/116 & 797/100 \\
		 \midrule
		 Maximum powered surface deformation (PV/RMS, in nm) & 586/113 & 1382/115 \\
		 \midrule
		 Flat map deformation (PV/RMS, in nm) & 89/3.4 & 787/17\\
		\bottomrule
	\end{tabular}
	\label{table:Surface_properties}
\end{table}

    \subsection{Electrical connections}
\begin{figure}[t]
    \centering
	\includegraphics[width=0.7\linewidth]{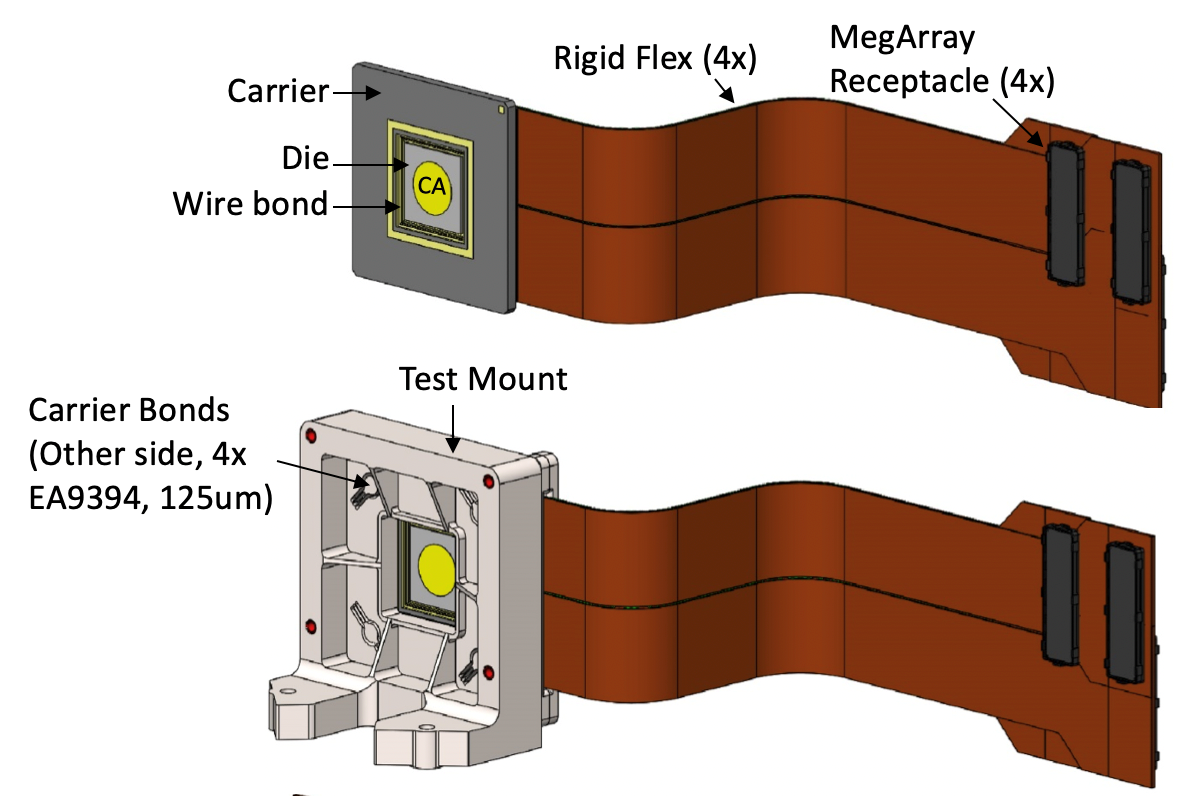}
	\caption{Front schematic of the ceramic chip carrier, mount, and rigid flex cables designed for the 2K MEMS DMs.} 
	\label{fig:Ceramic} 
\end{figure}

\begin{figure}[t]
    \centering
	\includegraphics[width=\linewidth]{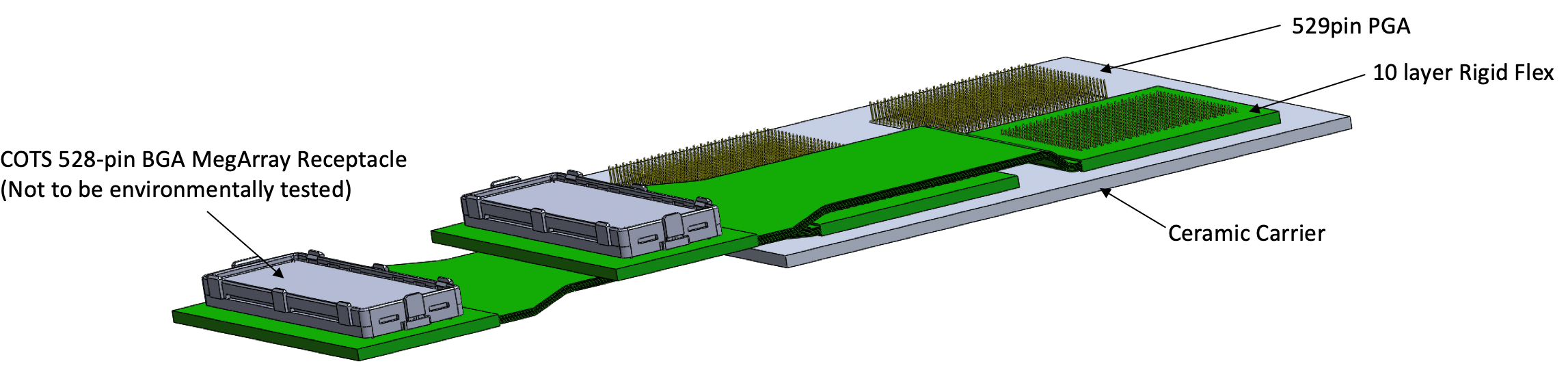}
	\caption{Pin-grid array (PGA) assembly at the back of the ceramic chip. The MEG-Array connector was not considered part of the random vibe test.} 
	\label{fig:MegArray} 
\end{figure}

For both DUTs, coated-dies were attached with adhesive to a ceramic package specifically designed for the 2K DMs (see in Fig.~\ref{fig:Ceramic}). The DM die and the ceramic package were electrically connected using gold wire-bonds applied with a high-precision automated tool at BMC’s facility. JPL also fabricated flex cables that were connected in the back of the new chip carrier through a pin-grid array (PGA) and terminated in 528-pin MEG-Array connectors (see Fig.~\ref{fig:MegArray}).

The packaged DMs were tested using high-voltage drivers commercially available from BMC to characterize their electro-mechanical and optical performance. BMC's electronics connect to the DM via the MEG-Array connectors. 
    
\section{Procedure}
\label{sec:Environmental_Testing}
    \subsection{Testing overview}
    \label{subsec:Probing}
\begin{figure}[t]
    \centering
	\includegraphics[width=\linewidth]{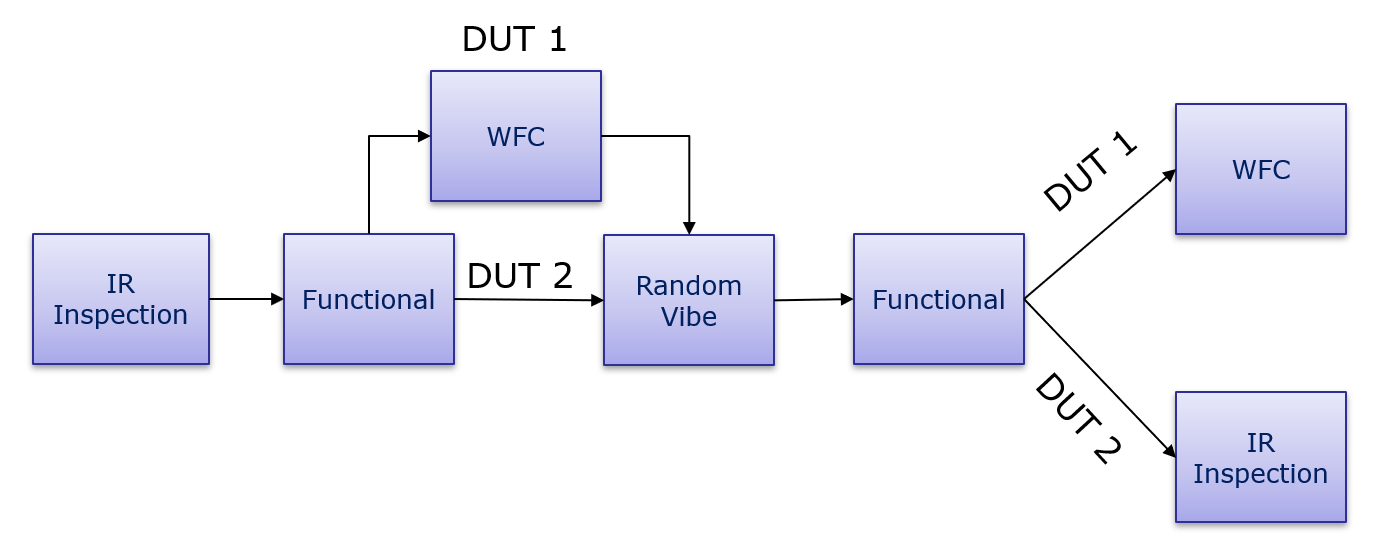}
	\caption{Testing workflow. The tests before exposure to random vibrations consisted of an IR microscope inspection and functional tests using a Fizeau interferometer. DUT~1 was also used for wavefront control (WFC) performance testing on a coronagraph testbed. After random vibe, each DM was functionally testing using an interferometer. DUT~1 was then used for WFC performance testing and DUT~2 was inspected using the IR microscope.} 
	\label{fig:Workflow_overview} 
\end{figure}

Both DUT 1 and 2 underwent a battery of tests before and after exposure to random vibrations. The carrier, die, die bonds, PGA joints and carrier to test mount bonds were inspected in all phases. The actuator functionality and performance were tested using the steps outlined below. The MEG-Array connectors and receptacles are not envisioned as part of the flight system and are therefore not included in our analysis. The workflow of these experiments for both DUTs is illustrated in Fig.~\ref{fig:Workflow_overview}. Results of these tests are described in Sec.~\ref{sec:Results}.

        \subsection{Infrared inspection}
        \label{subsubsec:IR_imaging}
Before applying the face sheet coating to DUT~1, both DUTs were inspected at BMC using transmissive infrared microscopy. The system was automated to translate the DM and image many actuators in sequence. To achieve this, a MLS203 fast X-Y stage was installed on a BX51 Olympus microscope. The microscope was also equipped with a MFC1 Motorized Microscope focus controller to focus either on the wiring or on the mirror layer. The wiring layer was inspected on the entire die of dimension 32.8$\times$32.8~mm by steps of 400~$\mu$m (size of the actuator pitch) for a total of 6,224 images. The mirror layer inspection was restricted to the device area to image each actuator individually and 4,080 images were taken in a serpentine clockwise path. In total, 10,804 images were recorded per DM, to be compared before and after random vibe in case of failure. DUT 2 had three actuators with clearly visible defects just after fabrication (see the infrared image of an anomalous actuator in Fig.~\ref{fig:Result_IR}).

        \subsection{Functional testing}
        \label{subsubsec:Zygo}
DUT~1 was then coated and both DMs were sent to the HCIT facility at JPL for testing. Functional testing was done using a Fizeau interferometer (Zygo Verifire). The DMs were placed inside of a plastic enclosure that was purged with a continuous flow of dry air to maintain a relative humidity of $<$30\% during operation and to avoid any electrostatic discharge event\cite{Morzinski2012}. The devices were connected through the MEG-Array connectors to a commercial 14-bit electronics provided by BMC for the 2K-DM to perform a battery of functional spatio-temporal tests. The input voltage was limited to 90~V. 

The HCIT team developed a series of functionality tests aimed at highlighting defective actuators before and after random vibe that were sorted in several categories. A ``pinned" actuator is fixed to its unpowered position and is easily noticeable when uniform voltage is applied to the DM. A ``free-floating" actuator does not move through electric commands but remains free to move to follow the displacements of its neighbors. ``Tied" actuators occur when more than one actuator respond to a command sent to a single actuator index. A ``weak" actuator moves significantly less than its neighbors with the same command. Finally, an ``anomalous" actuator can be any of the above categories or otherwise defective. The standard functionality test process consists of the routines described below. For each surface measurement recorded by the Zygo interferometer, the uncommanded surface was subtracted and piston, tip and tilt aberrations were removed from the data in post-processing.

The functionality test routines are as follows: 

\paragraph{Applying a uniform voltage.} A uniform voltage is applied to all actuators. The measurement is taken at increasing voltage levels. Pinned actuators are particularly apparent in the resulting data.

\paragraph{Poking individual rows and columns.} A uniform voltage is applied to the actuators in the same rows and then columns (100 measurements for a 2K MEMS DM). This highlights anomalous actuators and helps determine their index. This is also used to confirm the mapping between high voltage channels in the electronics and the actuators. Tied actuators are also noticeable in the data, except if they are located in the same row or column. 

\paragraph{Poking grids of actuators.} The actuators are divided in $4\times4$ regions and one actuator of each region is poked simultaneously such that it creates a regular grid, with large enough separation to avoid coupling effects. This process is repeated 16 times to cover all the DM actuators. Free-floating actuators are particularly visible in the resulting data. This is a standard calibration routine for our DMs since it can be used to estimate the voltage to surface displacement conversion for each actuator using a limited number of images. 

\paragraph{Poking individual actuators.} The actuators are poked one-by-one. This is used to find the index of each anomalous actuator that was noticed in earlier stages and to solve ambiguities.

\paragraph{Stability measurement.} First, a uniform voltage is applied once to all actuators and one measurement is recorded every minute for two hours (uncommanded stability). Second, a uniform voltage is applied every minute for two hours immediately followed by a measurement with the interferometer (commanded stability). These tests aim to measure DM drifts over time. The low order spatial aberrations are filtered to monitor the drift of individual actuators. The mean of the recorded time series is subtracted for each image and the standard deviation is measured. An animation of the processed images is also visually inspected for anomalies.

\paragraph{Repeatability measurement.} The voltage is cycled between zero and a uniform voltage for all actuators every five seconds for 50~s and a measurement is recorded after each cycle. This aims to highlight any hysteresis due to the DM or the electronics. The images are processed the same way as the stability measurements.

\paragraph{Temporal response measurement.} We apply zero volts to the DM followed by a uniform bias. Ten measurements are then recorded as quickly as possible with the interferometer for about 20~s. The aim is to identify slow responding actuators. One measurement takes about two second on average which prevents detection of temporal frequencies higher than 0.25~Hz. The low order spatial aberrations are removed in post-processing and the time series is subtracted by its last image to highlight any differences in our visual inspection. 

\paragraph{Calibrating the DM.} The grid of actuators is used to determine the locations of each actuators with respect to the Zygo beam and the surface displacement for a given voltage. The DM is then flattened iteratively using the information collected in this way. The linear or quadratic voltage-to-surface height conversion is then measured at flat DM state, which is then used in the model for the wavefront sensing and control method used to create the dark hole in the coronagraph instrument.

\begin{figure}[t]
    \centering
	\includegraphics[width=\linewidth]{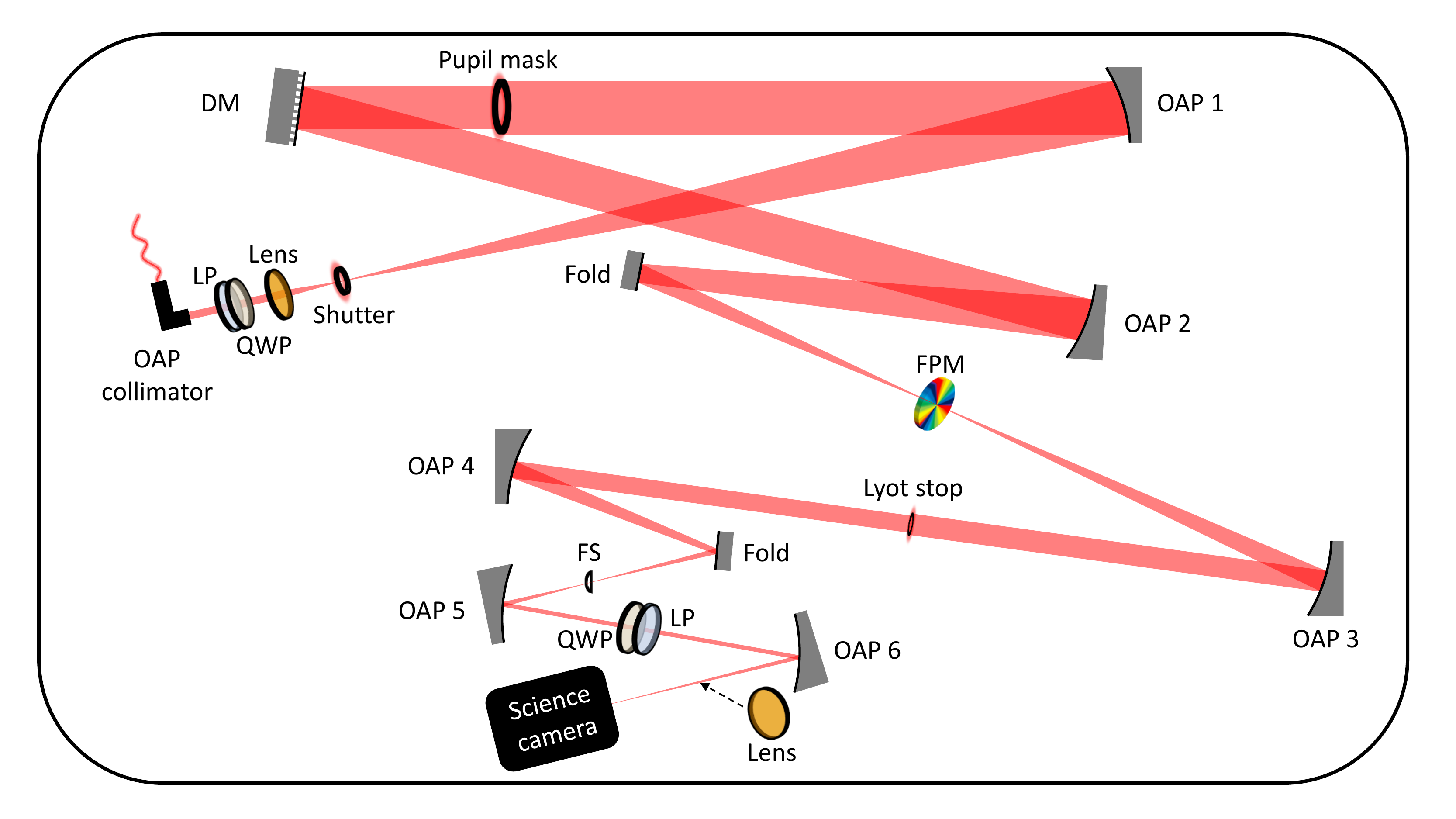}
	\caption{Schematic of the optical layout of the In-Air Coronagraph Testbed (IACT) in the HCIT facility at JPL. Not to scale.} 
	\label{fig:IACT_Layout} 
\end{figure}  

\subsection{Performance testing}
\label{subsubsec:IACT}

DUT~1 was also tested for high-contrast imaging purposes using the In-Air Coronagraph Testbed (IACT)\cite{Baxter2021} in the HCIT facility at JPL. The IACT optical layout is shown in Fig.~\ref{fig:IACT_Layout}. A 637~nm monochromatic light source was injected into the enclosed testbed through a single mode fiber to simulate a star. A charge 6 vector vortex coronagraph (VVC) was used to limit the sensitivity to low order wavefront aberrations due to air turbulence\cite{Foo2005,Mawet2009,Mawet2010,Ruane2018a}. 

The injected light was passed through a linear polarizer (LP) and a quarter wave plate (QWP) to circularly polarize the light source upstream of the focal plane mask (FPM). The LP had an extension ratio of $10^5$ and the QWP had a retardance of 0.24$\lambda$ at 637nm. The off-axis parabola (OAP) 1 then collimated the beam and reflected it toward a 18.48~mm pupil, immediately followed by DUT~1. There were 46.2 actuators across the beam at the DM. Similar to the functional test described above, DUT~1 was placed inside a plastic box with a continuous flow of dry air to reduce the humidity. To limit air turbulence, the flow was optimized to reach a relative humidity of 25\% to meet the DM specification. A Fluke DewK thermo-hygrometer was inserted in the dry box to actively sense the humidity level through a software watchdog. The box had an opening on the front side to allow the beam to reflect off DM to the 1524~mm focal length OAP2 that focused the light on the VVC FPM. The FPM was fixed to a 3-axis mount. The diffracted light was then reflected on the 762~mm focal plane OAP3 and blocked by a Lyot Stop (LS) of diameter 7.5~mm on a 2-axis mount. Considering the magnification of the OAPs, the LS diameter was 81.2\% of the pupil image. 

A ``D" shaped field stop (FS) of size 3-10$\lambda/D$ was added in the downstream conjugated focal plane. The purposed of the FS was to enhance the contrast in the final focal plane at the camera by minimizing stray light or photo-electrons inside the corrected regions adjacent to saturated regions. The FS was placed on a 3-axis mount to optimize focus and can be moved in and out for calibration purposes. The dark images that are later subtracted from the dark hole images were recorded by fully blocking the light at the FS plane. 

After the FS, the beam is then collimated by OAP5 to pass through another set of QWP and LP that minimize the incoherent leakage caused by the imperfect retardance in the VVC FPM. The rotation angles of the QWP+LP were optimized by minimizing the signal on the science detector with the VVC FPM fully removed from the beam. Finally, the OAP6 directed the light to the science detector where the final image is formed. A neutral density filter wheel can be used for calibration purposes, for instance to prevent the over-exposure of the unocculted PSF used for calibration, and also has an optical lens to allow pupil imaging. The science camera Andor Neo sCMOS electrically cooled to -40~\degree C and the generated heat was removed with a water cooler. The camera was mounted on a single-axis stage to control the focus. The pixel pitch was 6.5~$\mu m$ and the resolution of the focal plane images is 24.7 pixels per $\lambda/D$.

Standard wavefront sensing and control (WS\&C) algorithms and calibration procedures dedicated to high-contrast imaging were used to minimize the simulated stellar intensity at the detector plane and improve the raw contrast level (intensity of the attenuated starlight normalized by the maximum of the unocculted PSF)\cite{Ruane2022}. Phase retrieval algorithms based on both Gerchberg–Saxton formalism\cite{Gonsalves1982} and the fitting of low-order Zernike modes were used to flatten the DM and calibrate its response. At least three images close to the focal plane and three images close to the pupil plane were used to run the algorithm. After flattening the DM the Strehl ratio of the unocculted PSF was very close to 1.0. The VVC FPM and the LS were automatically centered on the beam iteratively through the acquisition of pupil images. Dark images and off-axis PSFs were then recorded and the FS is introduced in the beam to allow the desired off-axis dark hole region to pass through. 

Wavefront sensing and wavefront control are performed respectively with pair-wise probing (PWP) and electric field conjugation (EFC)\cite{Borde2006,GiveOn2007SPIE} through FALCO software\cite{Riggs2018}. Both algorithms require a high-performance DM to achieve contrast levels of $\sim10^{-8}$. The dark hole (DH) region where the stellar residuals are attenuated is defined by the FS aperture that goes from 3 to 10~$\lambda/D$. While the contrast improves in the DH, the exposure time on the science camera is increased from 100~s to 300~s to maintain a sufficient signal-to-noise ratio. The $\beta$-bumping technique\cite{Sidick2017} is regularly used to achieve the best possible coherent contrast in the dark hole region. The raw contrast in the dark hole is intricately linked to the DM performance. Performance testing results for DUT1 are presented in Sec.~\ref{subsubsec:DUT1}.

\begin{figure}[t]
    \centering
	\includegraphics[width=12cm]{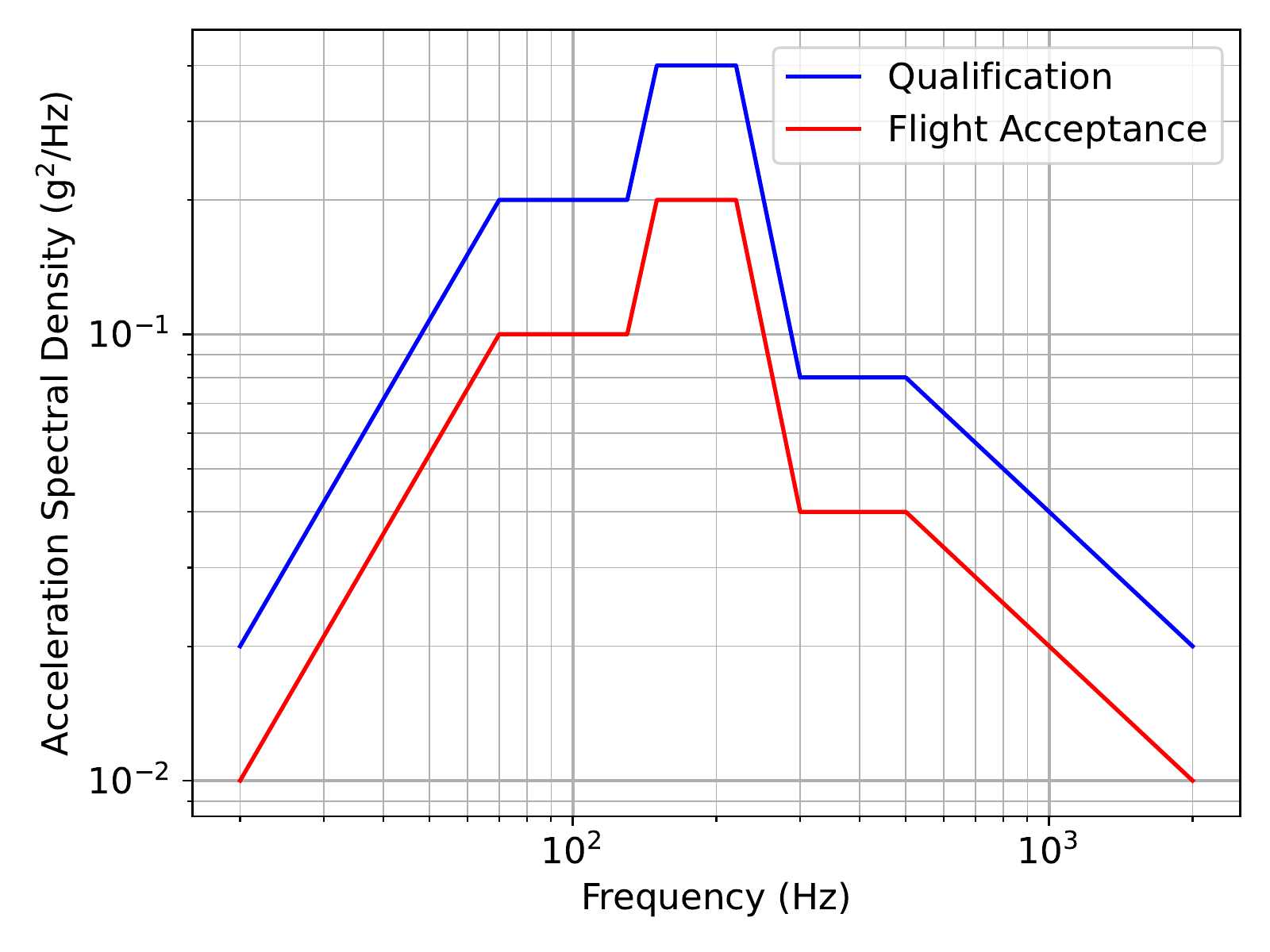}
	\caption{Power spectral density of the vibration experienced by the MEMS DMs in the three axes.} 
	\label{fig:PSD_Vibrations} 
\end{figure}

        \subsection{Random vibration environment}
        \label{subsec:Simulating_space}
        
In previous work, we have demonstrated robustness of actuators that were surrounded by functional actuators under flight-like thermal cycles and vibrations as well as the compatibility of partially-functional 50$\times$50 MEMS DMs with vacuum environments\cite{MejiaPrada2021}. This work expands the previous study since we specifically test the robustness of actuators at the vicinity of defective actuators as well as fully-functional 50x50 MEMS DMs.. The DMs and their respective mounts were shaken at JPL on a 10-inch cube shaker. The flex cable was fixed to the edge of the mount with a flex clamp while the other end of the flex was curled loosely and taped down onto the moving platform of the shaker to avoid any damage. Particle contamination and humidity were controlled during the test to ensure the DMs are not subject to alternative sources of electric degradation and failure\cite{Huang2012}, as suspected in previous studies\cite{Bierden2022}. The applied signal ranges between 20~Hz and 2000~Hz, which corresponds to typical frequency ranges for most launch vehicles. The temporal acceleration spectral density of the vibrations in each of the three spatial axes was between 0.01~g$^2$/Hz and 0.4~g$^2$/Hz and is shown in Fig.~\ref{fig:PSD_Vibrations}. The DMs therefore underwent 11.7~gRMS over all frequencies for 2~min per axis. This qualification test is conservative to any potential launch vehicles and surpasses the flight acceptance followed by the Roman Space Telescope Coronagraph Instrument specifications.

\begin{figure}[t]
    \centering
	\includegraphics[width=\linewidth]{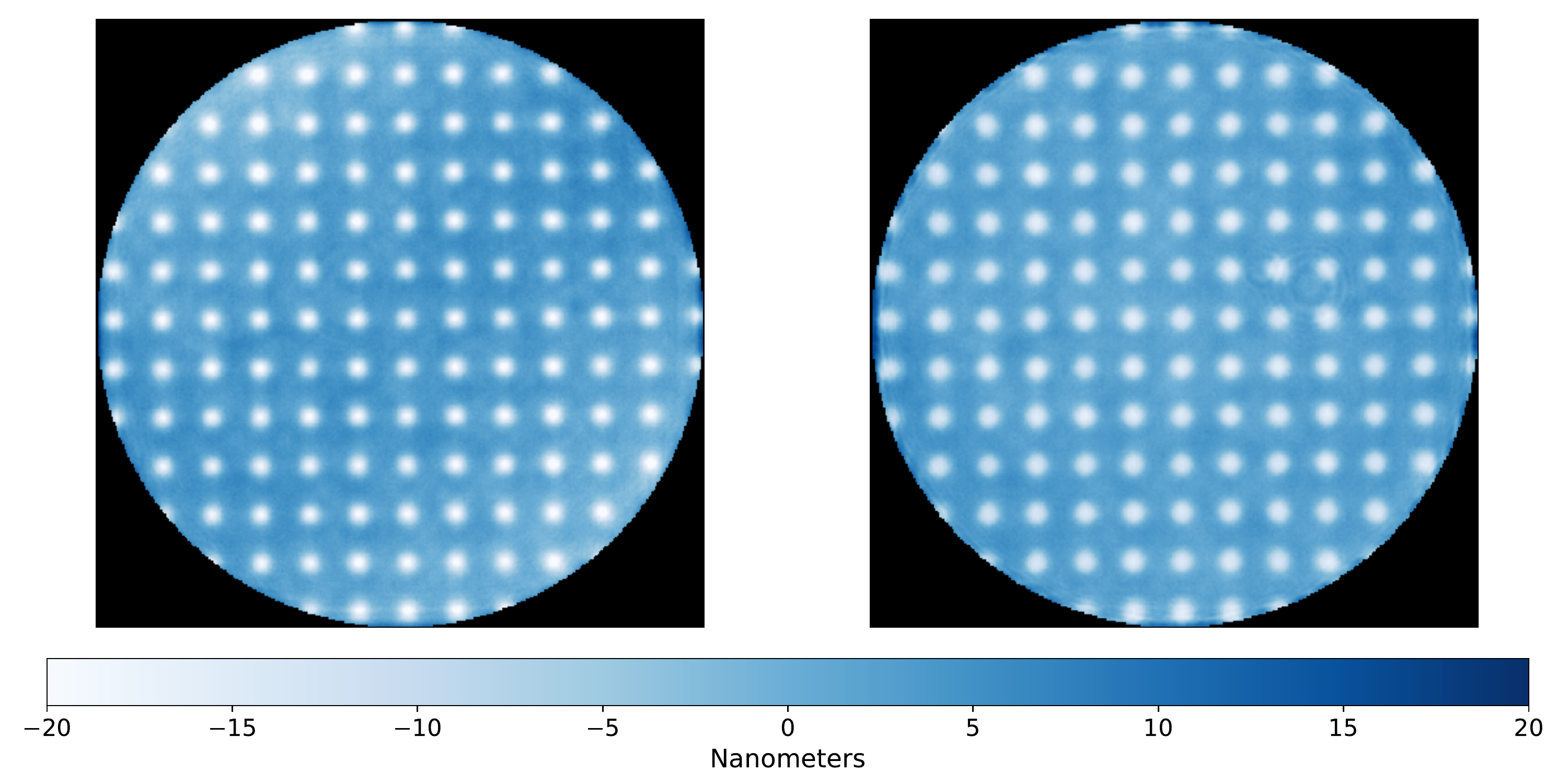}
	\caption{Optical path difference (in nanometers) that corresponds to a functional test where a regular grid of actuator was poked, before (left) and after (right) the shaking of DUT1.} 
	\label{fig:Result_FunctionalF} 
\end{figure}  


\section{Test results}
\label{sec:Results}

\subsection{DUT 1}
\label{subsubsec:DUT1}

DUT1 is a 100\% yield device intended to validate that a fully functioning MEMS DM passes random vibration testing. The criteria of success are visual inspection of the structure; the actuator responsiveness, stroke, and voltage-to-displacement gain; and the ability to create a dark hole at relevant contrast levels ($\sim10^{-8}$). This section compares the results of functional and performance testing before and after random vibration at JPL.

Fig.~\ref{fig:Result_FunctionalF} represents one of the 16 regular grids of actuators that were poked during the functional test of DUT1 before and after the random vibe test. The image resolution has been slightly decreased after vibe because we did not insure same Zygo zoom settings before and after random vibe test. Such a resolution remains acceptable for direct comparison with pre-vibe data since we have many more than the required 4 pixels per DM actuator. As in the 15 other grids, the behavior of the actuators was identical before and after random vibe regardless the applied voltage. No anomaly in the influence function shape nor displacement of the DM surface occurred. After DM flattening, the final shape was measured to be 3.41~nm RMS wavefront error before and 3.37~nm RMS after random vibe. Around the flat setting, the quadratic relationship between the surface displacement amplitude of each actuator and the applied voltage remains identical before and after vibe.
All functional tests performed did not show any anomalies on DUT1 either before and after the DM was random vibed.

\begin{figure}[t]
    \centering
	\includegraphics[width=\linewidth]{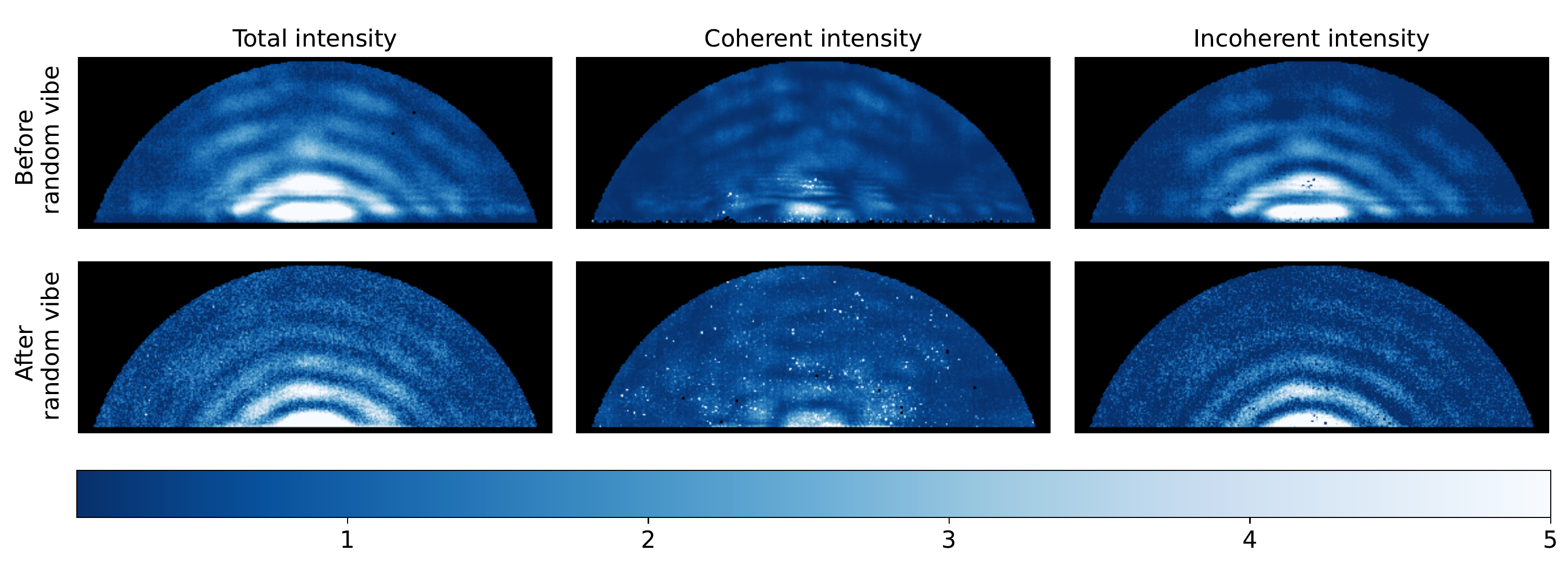}
	\caption{Post-WS\&C contrast maps ($\times10^{8}$) before (top) and after (bottom) the DUT1 underwent flight-like shaking. The total (left), coherent (center) and incoherent (right) intensities are presented. The exposure time for both total intensity images is 300~s but the source injection unit has been moved between the experiments, explaining the noise discrepancy.} 
	\label{fig:Result_Performance} 
\end{figure}  
After functional testing, DUT1 was installed on IACT to test the DM performance in a coronagraph instrument. Figure~\ref{fig:Result_Performance} shows the normalized intensities in the science image before and after random vibe, and after a few dozen WS\&C iterations. The mean contrast in the DH is $1.19\times10^{-8}$ (before random vibe) and $9.53\times10^{-9}$ (after random vibe) while the spatial standard deviation is equal to $1.42\times10^{-8}$ and $1.28\times10^{-8}$, respectively. The mean coherent contrast is equal to $3.80\times10^{-9}$ before and $4.34\times10^{-9}$ after with a respective standard deviation of $3.71\times10^{-9}$ and $3.86\times10^{-9}$. The small difference in coherent contrast is explained by a slightly higher internal turbulence on the testbed after random vibe. We also observe in Fig.~\ref{fig:Result_Performance} some horizontal artifacts that result from diffraction effects caused by a slight misalignment of the field stop in the z-direction. These highly localized effects did not impact convergence of the PW+EFC algorithm nor the computation of contrast performance. The decrease of flux after random vibe might induce an underestimation of the mean incoherent intensity leading to a slight overestimation of the contrast performance after random vibe.
\begin{figure}[t]
    \centering
	\includegraphics[width=12cm]{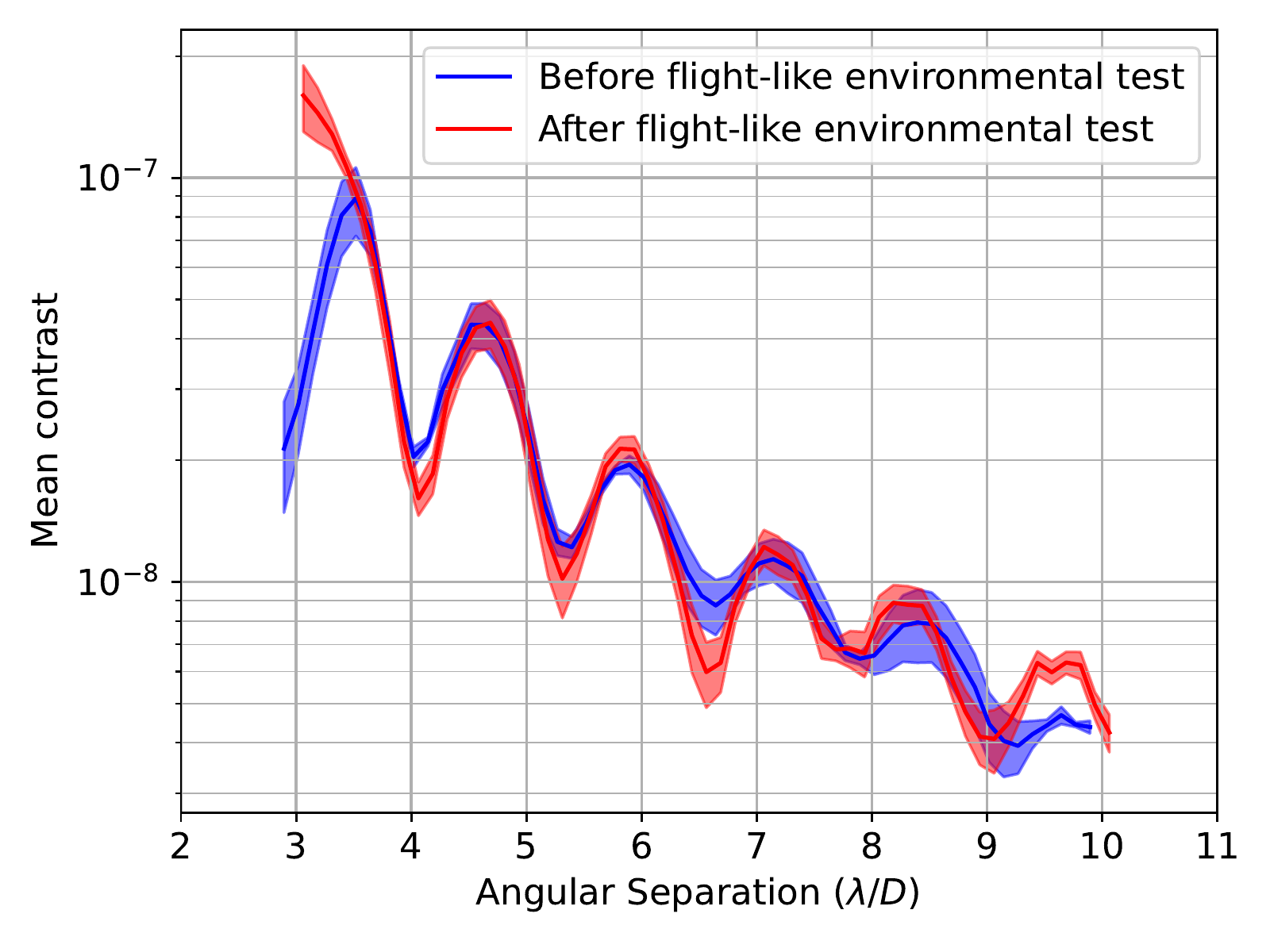}
	\caption{Post-WS\&C radial profiles of the raw contrast on the science detector before (blue) and after (red) the DUT1 underwent random vibe testing.} 
	\label{fig:Result_Performance_Contrast} 
\end{figure}  

Figure~\ref{fig:Result_Performance_Contrast} overlays the radial profile of each total intensity image whose mean contrast is calculated in annulus of $\lambda/8D$. On one hand, both Fig.~\ref{fig:Result_Performance} and Fig.~\ref{fig:Result_Performance_Contrast} emphasize that IACT performance are limited at low separations by an Airy pattern. This pattern is not sensed by PWP. This pattern is known to be an incoherent leakage due to manufacturing defects in the VVC FPM and in the LP and QWP\cite{Ruane2022}. This leakage could be further reduced by improving the retardance error in the QWP and extinction of the LP that are currently used. On the other hand, the coherent component in the DH was measured below $10^{-8}$ in both cases and its speckle intensity structure was modified at each iteration. We therefore attribute the remaining coherent component to the internal turbulence in IACT on timescales of a single WS\&C iteration. This effect could be reduced on IACT by lowering the dry air-flow or by installing an additional WS\&C system to specifically control low order spatial aberrations at higher temporal frequencies \cite{Shi2016}. Nonetheless, from the results of both performance and functional test on DUT1, we can conclude that DUT1, a 100\% functional 2K MEMS DM, survived random vibrations similar to a launch vehicle.

\subsection{DUT 2}
\label{subsubsec:DUT2}
\begin{figure}[t]
    \centering
	\includegraphics[height=10cm]{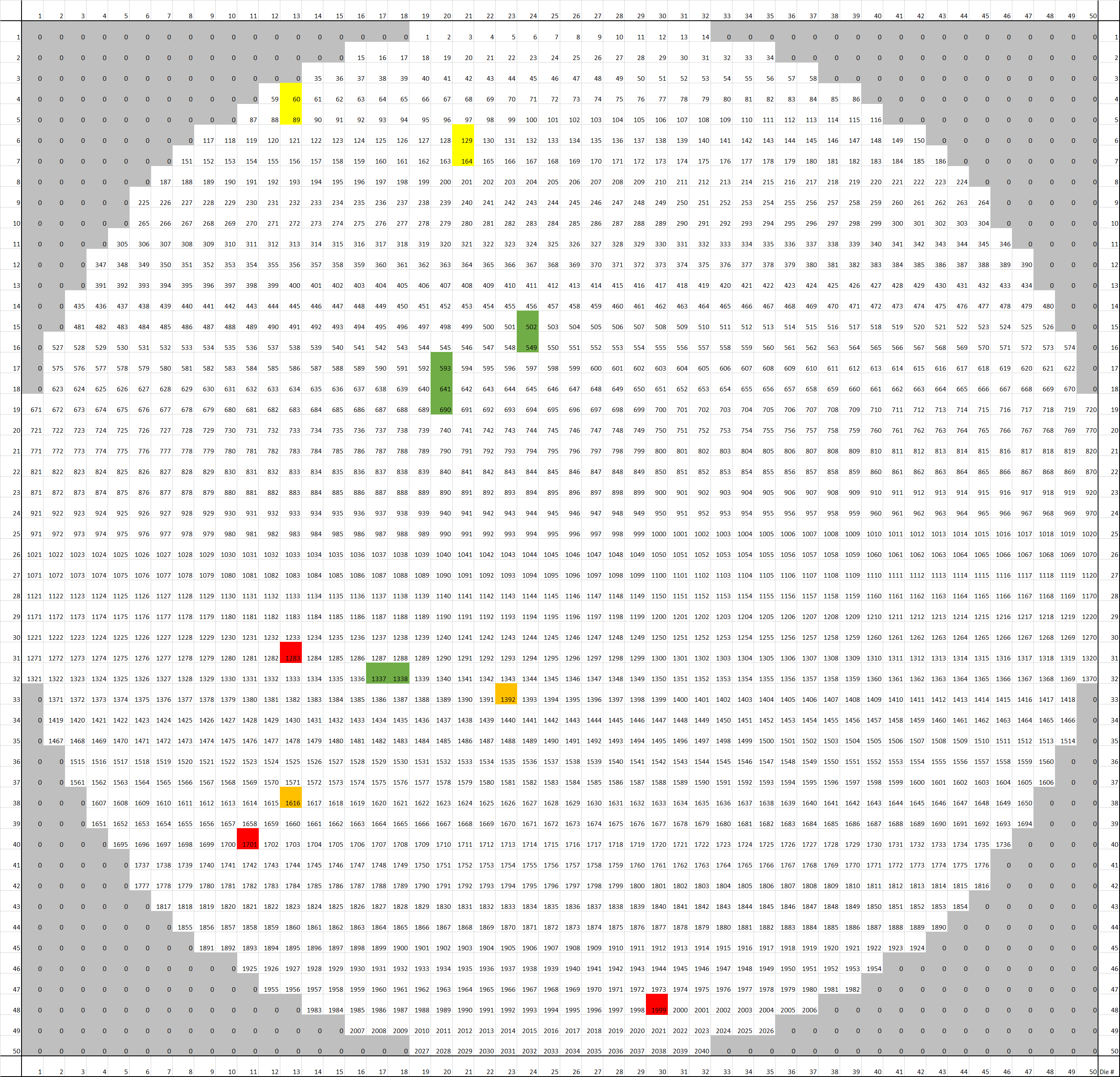}
	\caption{Grid of DUT2 actuators. Yellow: Tied actuators (60-89, 129-164). Green: Tied and weak actuators (502-549, 593-641-690, 1337-1338). Red: Pre-vibe pinned actuators (1283, 1701, 1999). Orange: Post-vibe pinned actuators due to poor connections at the MEG-Array connectors level (1392, 1616).} 
	\label{fig:Grid_DUT2} 
\end{figure}  

DUT2 had a few defective actuators and no metallic coating with the intention of testing the hypothesis that anomalous actuators can propagate to neighbors during rocket launch. DUT2 is not coated to allow IR inspection after random vibe. DUT2 also underwent the battery of functional tests described above, but was not used for performance testing.

\begin{figure}[t]
    \centering
	\includegraphics[width=\linewidth]{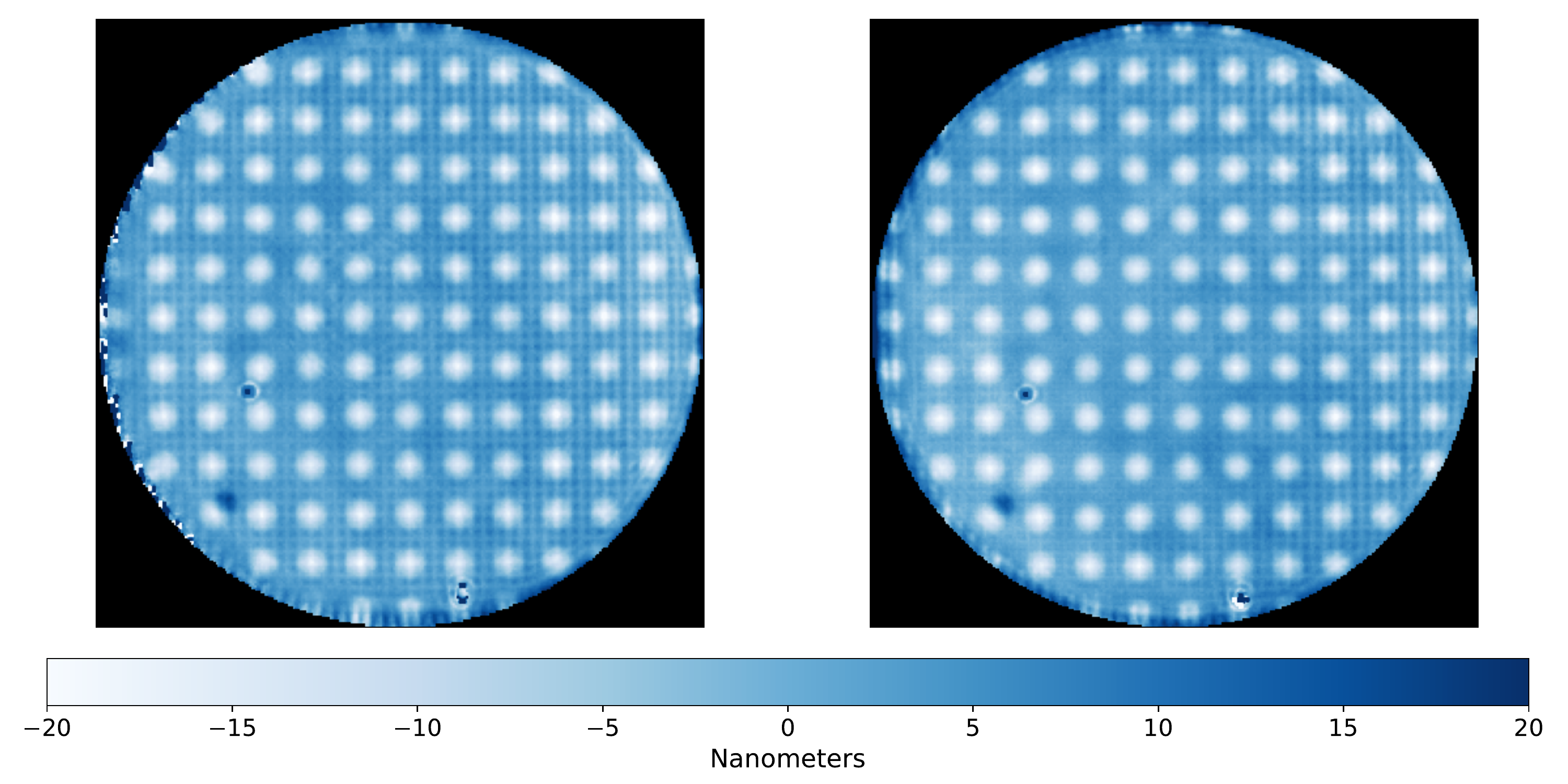}
	\caption{Optical path difference (in nanometers) that corresponds to a stage of functional test, where a regular grid of actuator was poked, before (left) and after (right) random vibe testing of DUT2.} 
	\label{fig:Result_FunctionalG} 
\end{figure}
\begin{figure}[t]
    \centering
	\includegraphics[width=10cm]{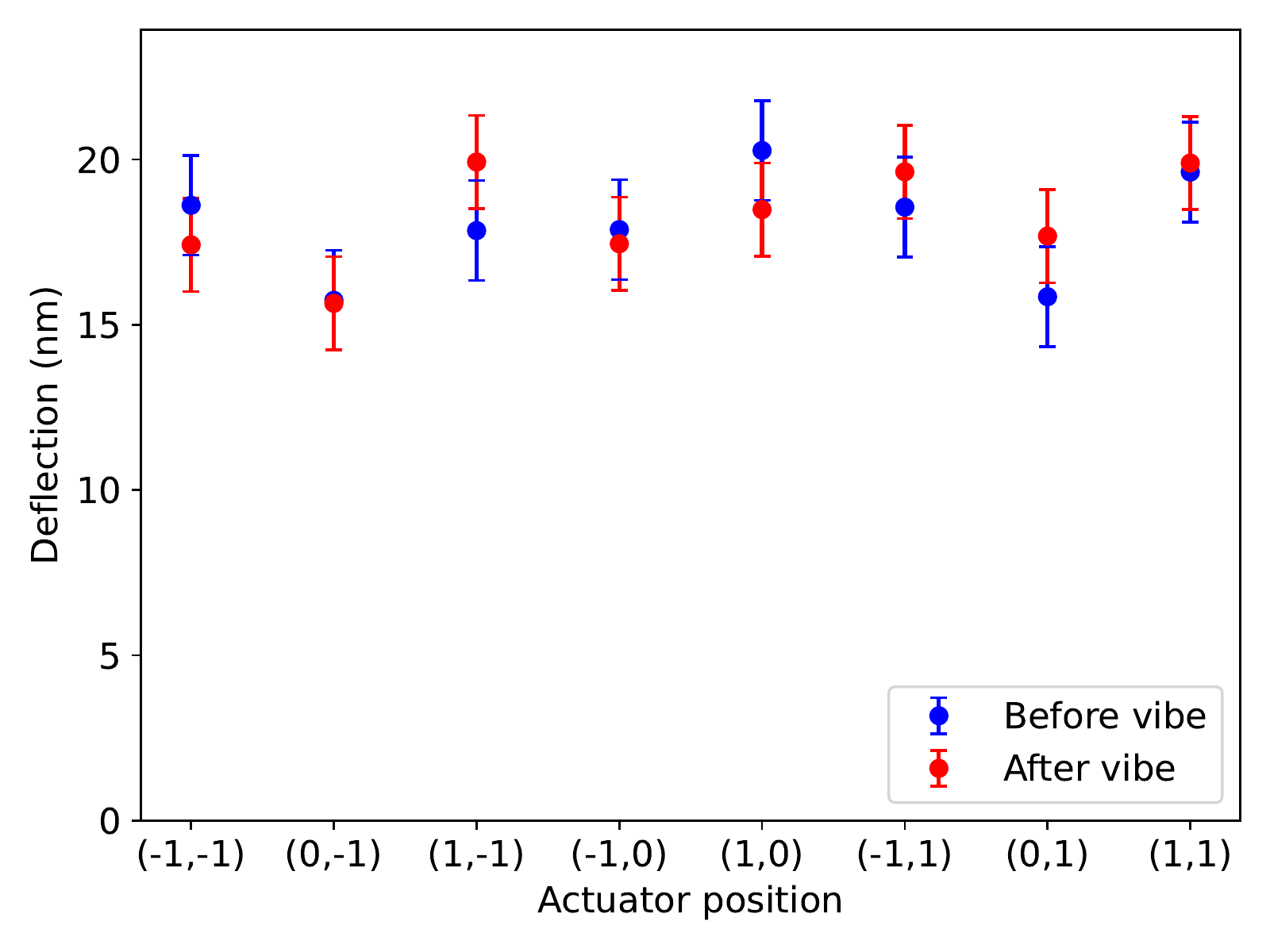}
	\caption{Deflection of actuator 1283's neighbors. We applied 0.025 BMC unit on these individual actuators on top of 0.05 BMC unit applied to all DUT~2 actuators.} 
	\label{fig:Deflection1283} 
\end{figure}

  
In pre-vibe function testing, DUT2 was found to be $\sim99.3$\% functional (see in Fig.~\ref{fig:Grid_DUT2}: it had three pinned actuators, two couples of tied actuators, and two couples and one triplet of weak and tied actuators (whose voltage to amplitude gain is divided by the number of associated actuators). One result of functional test is shown  in Fig.~\ref{fig:Result_FunctionalG} where the same grid of actuators was poked with respect to the grid in Fig.~\ref{fig:Result_FunctionalF}. The poke grid measurements show that DUT2 did not change behavior at actuator level. As an example, Fig.~\ref{fig:Deflection1283} shows the deflection for neighbors of one defective actuator (index 1283) while applying individual voltage of 0.025 BMC unit on top of a flat bias of 0.05 BMC unit. Their influence functions have been fitted with a Gaussian whose maximum amplitude is reported in this plot. Error bars are computed as the standard deviation of the measured amplitude for the 8 neighbors. Deflection of 1283 actuator's neighbors remain identical before and after random vibration: failure did not propagate during rocket launch simulation. These results were confirmed by the remaining functional tests, described in Sec.~\ref{subsubsec:Zygo}.

Preliminary tests with post-vibe DUT2 presented new anomalous actuators with respect to pre-vibe, particularly tied characteristics. After further investigation, we realized these anomalies were due to poor connections at the MEG-Array connectors level rather than the DM. Indeed, the connectors were disconnected before the random vibe and then reconnected for the functional tests. This process is sensitive since the pins can be easily bent if the connectors are not carefully handled. The defective connectors were fixed either by disconnecting and then reconnecting the faulty MEG-Array connector or by replacing the connector savers if the initial reconnection appeared unsuccessful. The state of the MEG-Array connectors was carefully inspected throughout the whole process. Given the challenges faced by our team related to connectors, we advocate for the development of more robust high-density connector technology and more practical DM driver electronics\cite{Bendek2020}.

\begin{figure}[t]
    \centering
	\includegraphics[width=12cm]{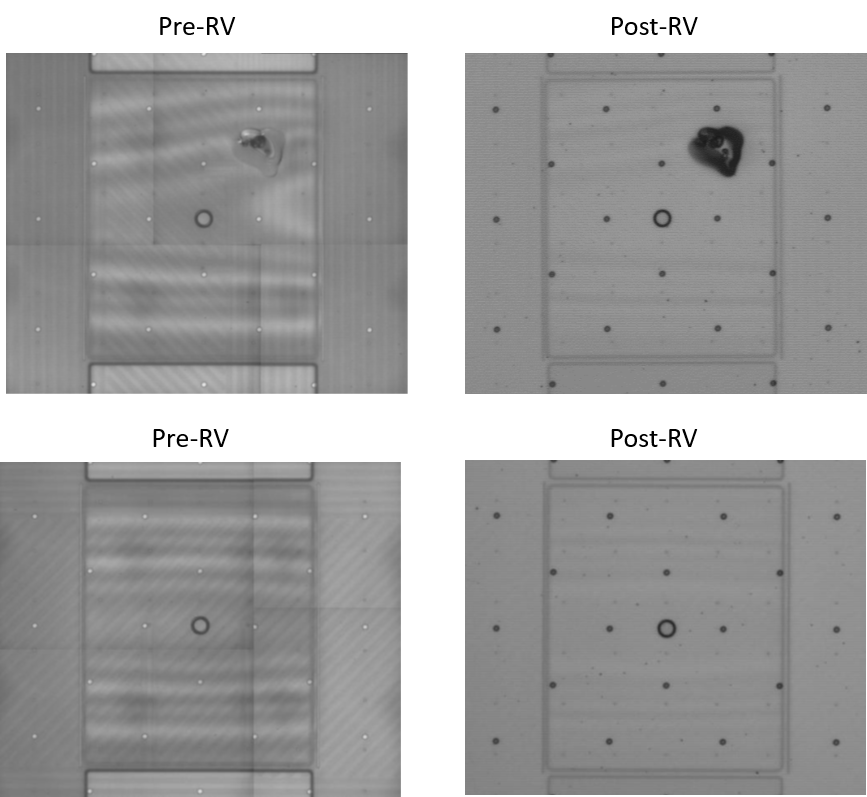}
	\caption{Pre- (left) and post- (right) vibe infrared images of both a pinned actuator (top) and the direct neighbor of a pinned actuator (bottom). Left images are focused on the wiring layer. Right images are focused on the mirror layer in reflection due to the DM package.} 
	\label{fig:Result_IR} 
\end{figure}

We also imaged the initial defective actuators and their neighbors with the infrared microscope to confirm the DM was not affected by the simulated rocket launch. No changes to the anomalous actuators nor their neighbors were notice during the post-vibe infrared inspection. Figure~\ref{fig:Result_IR} shows the infrared image of one tied actuator as well as the neighbor of another, focusing either on the wiring layer or on the mirror layer, before and after random vibe. The anomaly shown on the pinned actuator is apparent on both layers. The comparison of the images before and after random vibe shows that the damage has not propagated from the initial defect. The second set of images shows that none of the neighbor carrier, die, die bonds, PGA joints, nor actuators were affected by the random vibe test. From these results, we saw no evidence that anomalous actuators propagate to neighbors during random vibe.

\section{Conclusion}
\label{sec:Conclusion}
As part of a NASA SBIR, BMC and JPL jointly developed a new fabrication process for 50$\times$50 MEMS DMs. Two of these DMs underwent a battery of experiments to test their ability to survive in a launch vehicle. We have demonstrated that 1) a 100\% functional 2K MEMS DM maintains 100\% functionality and 2) that anomalous actuators do not propagate to neighboring actuators after undergoing launch-level vibrations. In conclusion, BMC's 2K continuous face sheet MEMS DMs have passed three-axes random vibe environmental testing at bounding launch loads encompassing those of future launch vehicles. Acoustics, shock, and radiation testing remain key steps towards achieving TRL~6 for BMC's MEMS DM technology. In addition, we recommend further development of connector systems for flight DMs in order to lower the risk of creating anomalous actuators during future DM testing, flight qualification, and mission development.

\acknowledgments 
The research was carried out at the Jet Propulsion Laboratory, California Institute of Technology, under a contract with the National Aeronautics and Space Administration (80NM0018D0004). The authors thank the anonymous reviewers for their
detailed feedback of this manuscript.

\subsection*{Disclosure}
This paper is the end product of the intermediary work presented in the SPIE Proceedings: Prada et al. 2021 ("Environmental testing of high-actuator-count MEMS
deformable mirrors for space-based applications," Proc. SPIE 11823,
Techniques and Instrumentation for Detection of Exoplanets X, 118230M (1
September 2021); doi: 10.1117/12.2594263).


\bibliography{bib_GS}   

\begin{thebibliography}{10}

\bibitem{Astro2020}
{National Academies of Sciences, Engineering, and Medicine}, {\em Pathways to
  Discovery in Astronomy and Astrophysics for the 2020s}, The National
  Academies Press, Washington, DC  (2021).

\bibitem{Malbet1995}
F.~{Malbet}, J.~W. {Yu}, and M.~{Shao}, ``{High-Dynamic-Range Imaging Using a
  Deformable Mirror for Space Coronography},'' {\em Pub.\ Astron.\ Soc.\
  Pacific} {\bf 107}, 386  (1995).

\bibitem{Trauger2004}
J.~T. {Trauger}, C.~{Burrows}, B.~{Gordon}, {\em et~al.}, ``{Coronagraph
  contrast demonstrations with the high-contrast imaging testbed},'' {\em Proc.
  SPIE} {\bf 5487}, 1330--1336  (2004).

\bibitem{Ealey2004}
M.~A. {Ealey} and J.~T. {Trauger}, ``{High-density deformable mirrors to enable
  coronographic planet detection},'' {\em Proc. SPIE} {\bf 5166}, 172--179
  (2004).

\bibitem{Wirth2013}
A.~{Wirth}, J.~{Cavaco}, T.~{Bruno}, {\em et~al.}, ``{Deformable mirror
  technologies at AOA Xinetics},'' {\em Proc. SPIE} {\bf 8780}, 87800M  (2013).

\bibitem{Bifano2011}
T.~{Bifano}, ``{Adaptive imaging: MEMS deformable mirrors},'' {\em Nature
  Photonics} {\bf 5}, 21--23  (2011).

\bibitem{Morgan2019}
R.~{Morgan}, E.~{Douglas}, G.~W. {Allan}, {\em et~al.}, ``{MEMS Deformable
  Mirrors for Space-Based High-Contrast Imaging},'' {\em Micromachines} {\bf
  10}(6), 366  (2019).

\bibitem{Bendek2020}
E.~{Bendek}, G.~{Ruane}, C.~M. {Prada}, {\em et~al.}, ``{Microelectromechanical
  deformable mirror development for high-contrast imaging, part 1:
  miniaturized, flight-capable control electronics},'' {\em J. Astron. Telesc.
  Instrum. Syst.} {\bf 6}, 045001  (2020).

\bibitem{Kasdin2020}
N.~J. {Kasdin}, V.~P. {Bailey}, B.~{Mennesson}, {\em et~al.}, ``{The Nancy
  Grace Roman Space Telescope Coronagraph Instrument (CGI) technology
  demonstration},'' {\em Proc. SPIE} {\bf 11443}, 114431U  (2020).

\bibitem{Baudoz2018SPIE}
P.~{Baudoz}, R.~{Galicher}, A.~{Potier}, {\em et~al.}, ``{Optimization and
  performance of multi-deformable mirror correction on the THD2 bench},'' {\em
  Proc. SPIE} {\bf 10706}, 107062O  (2018).

\bibitem{MejiaPrada2019}
C.~{Mejia Prada}, E.~{Serabyn}, and F.~{Shi}, ``{High-contrast imaging
  stability using MEMS deformable mirror},'' {\em Proc. SPIE} {\bf 11117},
  1111709  (2019).

\bibitem{Riggs2021}
A.~J.~E. Riggs, G.~Ruane, C.~A. {Mejia Prada}, {\em et~al.}, ``{High contrast
  imaging with MEMS deformable mirrors in the Decadal Survey Testbed},'' {\em
  Proc. SPIE} {\bf 11823}, 118230S  (2021).

\bibitem{HabEx_finalReport}
B.~S. {Gaudi}, S.~{Seager}, B.~{Mennesson}, {\em et~al.}, ``{The Habitable
  Exoplanet Observatory (HabEx) Mission Concept Study Final Report}.''
  \url{https://www.jpl.nasa.gov/habex/pdf/HabEx-Final-Report-Public-Release.pdf}
   (2019).

\bibitem{LUVOIR_finalReport}
{The LUVOIR Team}, ``{The Large UV Optical Infrared Surveyor (LUVOIR) final
  report}.''
  \url{https://asd.gsfc.nasa.gov/luvoir/resources/docs/LUVOIR_FinalReport_2019-08-26.pdf}
   (2019).

\bibitem{Bierden2022}
P.~{Bierden}, ``{MEMS Deformable Mirror Technology Development for Space-Based
  Exoplanet Detection},'' {\em SAT Milestone report}   (2022).

\bibitem{MejiaPrada2021}
C.~{Mejia Prada}, D.~{Liu}, G.~{Ruane}, {\em et~al.}, ``{Environmental testing
  of high-actuator-count MEMS deformable mirrors for space-based
  applications},'' {\em Proc. SPIE} {\bf 11823}, 118230M  (2021).

\bibitem{Morzinski2012}
K.~M. {Morzinski}, A.~P. {Norton}, J.~W. {Evans}, {\em et~al.}, ``{MEMS
  practice: from the lab to the telescope},'' {\em Proc. SPIE} {\bf 8253},
  825304  (2012).

\bibitem{Baxter2021}
W.~{Baxter}, A.~{Potier}, G.~{Ruane}, {\em et~al.}, ``{Design and commissioning
  of an in-air coronagraph testbed in the HCIT facility at NASA's Jet
  Propulsion Laboratory},'' {\em Proc. SPIE} {\bf 11823}, 118231S  (2021).

\bibitem{Foo2005}
G.~{Foo}, D.~M. {Palacios}, and J.~{Swartzland er}, Grover~A., ``{Optical
  vortex coronagraph},'' {\em Opt. Letters} {\bf 30}(24), 3308--3310  (2005).

\bibitem{Mawet2009}
D.~{Mawet}, E.~{Serabyn}, K.~{Liewer}, {\em et~al.}, ``{Optical Vectorial
  Vortex Coronagraphs using Liquid Crystal Polymers: theory, manufacturing and
  laboratory demonstration},'' {\em Opt. Express} {\bf 17}(3), 1902--1918
  (2009).

\bibitem{Mawet2010}
D.~{Mawet}, L.~{Pueyo}, D.~{Moody}, {\em et~al.}, ``{The Vector Vortex
  Coronagraph: sensitivity to central obscuration, low-order aberrations,
  chromaticism, and polarization},'' {\em Proc. SPIE} {\bf 7739}, 773914
  (2010).

\bibitem{Ruane2018a}
G.~{Ruane}, D.~{Mawet}, B.~{Mennesson}, {\em et~al.}, ``{Vortex coronagraphs
  for the Habitable Exoplanet Imaging Mission concept: theoretical performance
  and telescope requirements},'' {\em J. Astron. Telesc. Instrum. Syst.} {\bf
  4}(1), 015004  (2018).

\bibitem{Ruane2022}
G.~Ruane, A.~J.~E. Riggs, E.~Serabyn, {\em et~al.}, ``{Broadband vector vortex
  coronagraph testing at NASA’s high contrast imaging testbed facility},''
  {\em Proc. SPIE} {\bf 12180}, 1218024  (2022).

\bibitem{Gonsalves1982}
R.~A. Gonsalves, ``{Phase Retrieval And Diversity In Adaptive Optics},'' {\em
  Opt. Eng.} {\bf 21}(5), 829--832  (1982).

\bibitem{Borde2006}
P.~J. {Bord{\'e}} and W.~A. {Traub}, ``{High-Contrast Imaging from Space:
  Speckle Nulling in a Low-Aberration Regime},'' {\em Astrophys.\ J.} {\bf
  638}, 488--498  (2006).

\bibitem{GiveOn2007SPIE}
A.~{Give'On}, B.~{Kern}, S.~{Shaklan}, {\em et~al.}, ``{Broadband wavefront
  correction algorithm for high-contrast imaging systems},'' {\em Proc. SPIE}
  {\bf 6691}, 66910A  (2007).

\bibitem{Riggs2018}
A.~{Riggs}, G.~{Ruane}, C.~T. {Coker}, {\em et~al.}, ``{Fast linearized
  coronagraph optimizer (FALCO) I: a software toolbox for rapid coronagraphic
  design and wavefront correction},'' {\em Proc. SPIE} {\bf 10698}, 106982V
  (2018).

\bibitem{Sidick2017}
E.~{Sidick}, B.-J. {Seo}, B.~{Kern}, {\em et~al.}, ``{Sensitivity of WFIRST
  coronagraph broadband contrast performance to DM actuator errors},'' {\em
  Proc. SPIE} {\bf 10400}, 1040006  (2017).

\bibitem{Huang2012}
Y.~Huang, A.~Vasan, R.~Doraiswami, {\em et~al.}, ``Mems reliability review,''
  {\em IEEE Transactions on Device and Materials Reliability - IEEE TRANS
  DEVICE MATER RELIA} {\bf 12}, 482--493  (2012).

\bibitem{Shi2016}
F.~{Shi}, K.~{Balasubramanian}, R.~{Hein}, {\em et~al.}, ``{Low-order wavefront
  sensing and control for WFIRST-AFTA coronagraph},'' {\em J. Astron. Telesc.
  Instrum. Syst.} {\bf 2}, 011021  (2016).

\end{thebibliography}
\bibliographystyle{spiejour}   


\listoffigures
\listoftables

\end{spacing}
\end{document}